# Conditions for diffusive penetration of plasmoids across magnetic barriers

N. Brenning, T. Hurtig, and M. A. Raadu

Alfvén Laboratory, Royal Institute of Technology, se-100 44 Stockholm, Sweden

**Abstract**

The penetration of plasma clouds, or plasmoids, across abrupt magnetic barriers (of the scale less than a few ion gyro radii, using the plasmoid directed velocity) is studied. The insight gained earlier, from experimental and computer simulation investigations of a case study, is generalised into other parameter regimes. It is concluded for what parameters a plasmoid should be expected to penetrate the magnetic barrier through self-polarization, penetrate through magnetic expulsion, or be rejected from the barrier. The scaling parameters are $n_e$, $v_0$, $B_{perp}$, $m_i$, $T_i$, and the width $w$ of the plasmoid. The scaling is based on a model for strongly driven, nonlinear magnetic field diffusion into a plasma, which is a generalization of the laboratory findings. The results are applied to experiments earlier reported in the literature, and also to the proposed application of impulsive penetration of plasmoids from the solar wind into the Earth's magnetosphere.

**1. Introduction.**

☐The subject investigated here is under what conditions fast plasma clouds (or plasmoids) can penetrate across abrupt magnetic barriers, which we here define as a situation where a transverse magnetic field components grows in space over less than a few ion gyro radii $r_{gi}$ (calculated using the transverse magnetic field component $B_{perp}$, and the plasma drift velocity $v_0$). We propose a model which can be used to determine, based on the parameters $n_e$, $v_0$, $B_{perp}$, $m_i$, $T_i$, and plasmoid width $w$, whether a given plasmoid will cross a magnetic barrier through self polarization ($\Delta B/B \approx 0$, and $\mathbf{E}_P \approx -\mathbf{v} \times \mathbf{B}$), cross through magnetic field expulsion ($\Delta B/B \approx 100\%$, and $\mathbf{E}_P \approx 0$), by a mixture of these processes, or become rejected from the barrier. The study is limited to the transition across the barrier, and does not include the further propagation in the region with the transverse magnetic field. There is a companion paper in the present issue [1], and a recent series of four papers from our group in Physics of Plasmas [2-5]. The present paper is mainly based on, and extends the work in [5].

In the companion paper [1] results were reported from a laboratory experiment. High frequency oscillations in the lower hybrid range were found to play a key role in the penetration process, allowing anomalous fast diffusion of the magnetic field into the plasma. Such fast magnetic diffusion is a key process in the penetration of the plasma across the magnetic barrier. In the present paper we will describe a generalization of the results from [1] into a model that can be applied in other parameter regimes. The paper is organized as follows. In section 2 we will give a brief review of four selected experiments with plasma penetration into transverse magnetic fields, which together cover a wide range of parameters. We will use data from these experiments to demonstrate four key experimental facts, in the form of open questions, which will subsequently be used to benchmark our model. In section 3 we will report further experiments in our plasma device and, based on them, make a generalization of the results in [1] to a model for strongly driven magnetic diffusion. We call it the nonlinear magnetic diffusion (NL-MD) model.  The NL-MD model is then demonstrated to be consistent with the four observational facts described in Section 2. Section 4 finally contains a discussion.

For clarity, we want to note here that this study is limited to investigate *diffusive* penetration in which waves, or turbulence, give anomalous fast magnetic diffusion. Alternative mechanisms, for example decoupling through parallel electric fields, have also been proposed, see [1]. A united investigation remains to be made of when one or the other (or none) is operating.

## 2. Background

Experiments have been made for several decades on plasma streams entering transverse magnetic fields, se references in [2-5]. We will here present data from four of these, chosen to cover the range of parameters of interest. We begin by describing these four experiments, and then give a brief overview of the main results and point at some open questions.

The experiments of Song *et al* [6] were made using a plasma source with high kinetic energy in the sense that $\beta_k > 1$ at the entry of the plasma into the transverse magnetic field, close to the source. Here $\beta_k = W_K/W_B$ is the ratio between the plasma's kinetic energy density and the transverse magnetic field energy density. All measurements were made at distances more than 2m from the plasma source. They are therefore made long after the magnetic transition that is our prime object of study, and after a phase of more than 2 m translation of the plasma across a transverse magnetic field. Considerable plasma expansion and sometimes 'post-transition' magnetic diffusion need to be kept in mind. We will here use data from two experiments, made at 50 Gauss (0,005T) and 300 Gauss (0,03T), for different purposes. In the 50 Gauss experiment there was almost complete magnetic expulsion maintained all the way the most distant measurement at 3,57m. This experiment here represents high normalized kinetic beam energy, $\beta_k > 1$, and penetration through magnetic expulsion. In the 300 Gauss experiment, the final stages of magnetic penetration were measured, from $\Delta B/B = 15\%$ at 2,05 m to $\Delta B/B = 2\%$ at 3,75m. At a distance of 2,69m from the source, the beam width was reported to be 18 times the ion gyro radius. The 300 Gauss experiment is here used to represent (using only the measurements at 2,69m) a plasmoid which propagates by self-polarization to $\mathbf{E} = - \mathbf{v}\times\mathbf{B}$, and with a large normalized width $w \gg r_{gi}$.

The experiments of Izishuka and Robertson [7] were made to investigate a proposed low energy density, 'electrostatic' limit for plasma stream penetration through self-polarization. The experiments were made using a charge-neutralized proton beam (140kV, 6-15A/cm$^2$, 800ns, from an ion diode), which was shot into a region with a transverse magnetic field of variable strength up to 0,3T. The normalized beam width was about unity, $w/r_{gi} \approx 1$. With an ion current of 12A/cm$^2$, corresponding to a plasma density of $1,4\times10^{17}$m$^{-3}$, they found a critical limit at a magnetic field strength of $B = 0,25$T. For lower magnetic field strengths, the plasma beam penetrated the magnetic barrier through self-polarization. For 0,25T and higher magnetic field strengths, the plasma beam failed to penetrate. This experiment here represents two phenomena, (1) the rapid magnetization and self-polarization of a low energy density beam, and (2) the 'electrostatic' lower limit to kinetic beam energy below which plasmoids can not penetrate a magnetic barrier.

The third experiment is made by our own group and is described in the companion paper [1] and in [3-5]. It represents a middle ground regarding most experimental parameters. Plasma streams had intermediate width, $w/r_{gi} \approx 0,3$ to 1, and normalized kinetic energy densities $\beta_k$ in the range 0,5 to 10. The plasma was, immediately after the transition, in a mixed state of magnetic field expulsion and self-polarization. The values of $\Delta B/B$ in the centre were (depending on the $\beta_k$ value) in the range 13% to 74%.

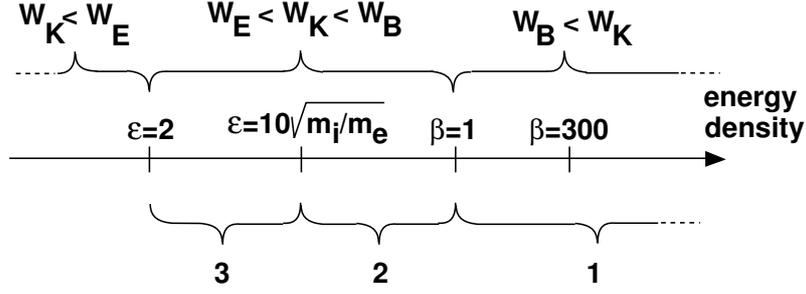

Fig. 1. A classification of plasma streams according to their energy density $W_K$. Above the axis are the dimensionless variables $\varepsilon = W_K/W_E$, and $\beta_k = W_K/W_B$. Below the axis are numbers 1 – 3 referring to three of the four questions listed in section 2.

The fourth experiment we will use is the Porcupine ionospheric injection experiment, as reported by Mishin *et al* [8]. The Porcupine experiment was made with a dense ($n_b \gg n_{amb}$) heavy ion (xenon, $m_i = 131$) beam, which was shot into the ambient ionosphere from a spinning rocket sub payload. Measurements were made on the main payload, at distances from 3 m when the beam was switched on, and out to distances of the gyro radius, 475 m. For a distance out to 4,5 m from the ion source there was a diamagnetic cavity ($\Delta B/B \approx 100\%$), in excellent agreement with theoretical expectations based on the source's known ion temperature of 50 eV (this cavity was maintained by transverse ion thermal pressure, which is a fundamentally different mechanism from the magnetic expulsion in the experiment by Song *et al* [6], and in our own experiment). Outside the magnetic cavity, the ambient magnetic field penetrated quickly, and almost completely, into the beam. Already at distances $r < 12\text{-}15$ m, $\Delta B/B \approx 0,2\%$ was measured. This experiment represents the self-polarization of a high-to-low energy density, narrow plasma beam ($w/r_{gi} \ll 1$), with very fast diffusion of the magnetic field.

These four experiments cover basic parameters from $\beta_k \ll 1$ to $\beta_k \gg 1$, from $w/r_{gi} \ll 1$ to $w/r_{gi} \gg 1$, and from magnetic expulsions $\Delta B/B$ from 0,2% to almost 100%. The experiment by Izhizuka and Robertson [7] pinpoints the low energy limit when the plasmoids cannot penetrate due to the electrostatic limit. Our own experiment covers a middle ground with respect to $\beta_k$ and $w/r_{gi}$, and is at the transition between self-polarization and magnetic expulsion. It also includes the most detailed investigation of the magnetic diffusion mechanism. Several questions arise from a consideration of how the material in these diverse experiments can be put in a single frame. We here consider four of these questions.

(1) Three important parameters are energy densities, as illustrated in Fig. 1: the directed kinetic energy density $W_K = m_i n v^2/2$ in the plasmoid, the electric field energy density $W_E = \varepsilon_0 E^2/2$ in a self-polarized plasmoid, and the magnetic field energy density $W_B = B^2_{perp}/2\mu_0$ of the transverse magnetic field in the magnetic barrier. Consider the parameter regime marked '1' in Fig. 1. A high energy ($\beta_k > 1$) plasmoid can always penetrate a magnetic barrier, either through magnetic field expulsion ($\Delta B/B \approx 100\%$, and $\mathbf{E}_P \approx 0$), through self-polarization ($\Delta B/B \approx 0$, and $\mathbf{E}_P \approx -\mathbf{v} \times \mathbf{B}$), or through a combination of both. However, the $\beta_k$ value alone does not determine what happens. In the experiment by Song *et al* [6] at 50 Gauss, in the experiment by ourselves [5], and in Porcupine [8] at 9 m from the source, we can find data that in all three cases correspond to values $\beta_k$ about unity. Yet the degree of magnetic expulsion ranges from less than 0,2% to almost 100%. Fig. 2 shows that the width of the plasmoid is an important parameter. A wide plasma stream tends to exclude the magnetic field, and a narrow plasma stream to become self-polarized. *Question 1 ($\beta_k$ >1)*: What determines if there is magnetic expulsion or self-polarization, and where is the limit?

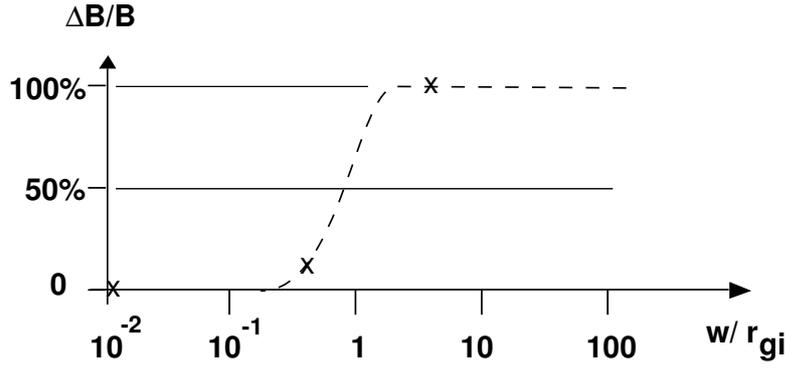

Fig. 2. The degree of magnetic expulsion in three experiments kinetic $\beta_k$ about unity, but with different normalized beam width $w/r_{gi}$. From left to right: Porcupine [8], the experiment by Hurtig *et al* [5], and the experiment of Song *et al* [6]. Numbers are given in Table I. The dashed curve is schematically drawn to represent an unknown relation.

(2) Consider the parameter regime marked '2' in Fig. 1. A low energy ($\beta_k$ <1) plasmoid has only energy enough to expel a fraction of the magnetic field, $\Delta B/B$ <1. It can therefore only penetrate through magnetic diffusion combined with self-polarization. The energy budget however requires that $\Delta B/B$ <1 is maintained at all times, also during the transition. The magnetic field has to penetrate the plasma on the same time scale as the transition time. For an abrupt transition in the sense studied here, this requires very efficient magnetic penetration, on the ion gyro time scale. Such mechanisms have to exist: the laboratory experiments by Ishizuka and Robertson [7], and also the Porcupine experiment [8] (for distances larger than 10 m from the source), show this for $\beta_k$ values far below unity. This requirement of fast magnetic diffusion becomes stricter for lower $\beta_k$, where there is only enough energy for a small fraction $\Delta B/B$ to be expelled. *Question 2 ($\beta_k$ < 1)*: What determines the residual value of $\Delta B/B$, and how is it energetically possible to create it for low $\beta_k$?

(3) Consider the parameter regime marked '3' in Fig. 1. In the lower end of this range there is a definite lower limit for penetration at energy density at $W_K = W_E$, when the kinetic energy is just sufficient to establish the self-polarization field. Theory [9] gives a higher limit, which has been called the electrostatic limit: $W_K \gg (m_i/m_e)^{1/2} W_E$. The exact value is uncertain. The only experiment where it has been determined [7] gave $W_K = 10(m_i/m_e)^{1/2} W_E$. *Question 3 ($\beta_k \ll 1$)*: What determines the low $\beta_k$, or electrostatic, limit for penetration?

(4) Lindberg [10] proposed an often quoted condition $w/r_{gi} < 0,5$ to set an upper limit to the width of self-polarized plasmoids, using an energy argument that is schematically illustrated in the upper panel of Fig. 3. In a self-polarized beam, there is in the laboratory rest frame a potential difference across the beam of $\Delta U = wE_P = v_0 w B_{perp}$. It is an experimental fact, also reproduced in simulations [2], that the low potential side of the beam is closest to the upstream potential, giving equipotential surfaces as indicated in Fig. 3. An ion that shall reach position (b) from the upstream plasma will therefore, argued Lindberg, in the transition have to climb a potential hill $\Delta U$. For this it needs to have a minimum energy $W_i > e\Delta U$, or $m_i v_0^2 / 2 > e v_0 w B_{perp}$. This is readily rewritten into Lindberg's condition $w/r_{gi} < 0,5$. On the other hand, in the experiment of Song *et al* [6] at 300 Gauss, there was a close to self-polarized beam ($\Delta B/B < 15\%$) with a width of $w/r_{gi} = 18$. This is illustrated in the lower panel of Fig. 3; the potential difference $\Delta U$ across the beam is here about 30 times the ion energy. *Question 4:* in what way is the penetration mechanism in the experiment by Song *et al* [6] different than envisaged by Lindberg [10]?

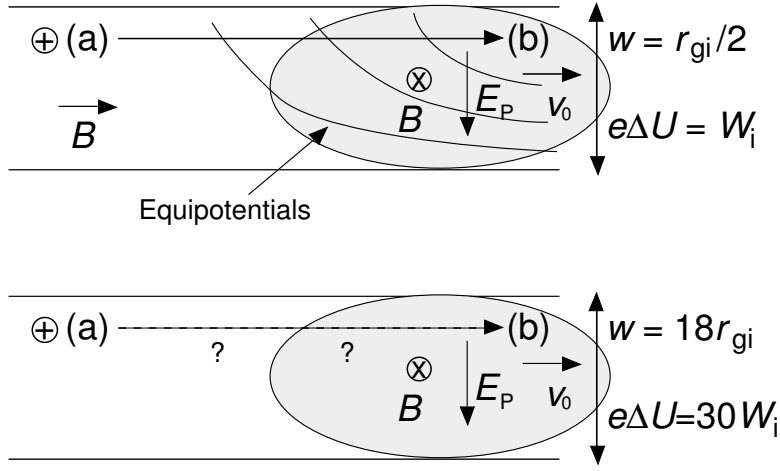

Fig. 3. Lindberg's condition (top panel) and the experiment by Song *et al* [6] at 300 Gauss (bottom panel).

## 3. The nonlinear magnetic diffusion (NL-MD) model

The waves, which were demonstrated in the companion paper [1] to produce anomalous fast magnetic diffusion into the plasma stream, were investigated closely in [3,4]. It was concluded that all wave features correspond well to a highly nonlinear version of the modified two-stream instability (MTSI), driven to high amplitude by a strong diamagnetic current $\mathbf{J}_D$. The high amplitudes reached during 1 µs of passage through the magnetic transition was consistent with this interpretation: linear theory for the MTSI [3] gives 10 – 26 linear growth times during the penetration time 1 µs. The effect of the waves was found [1,3,4] to be equivalent to an anomalous transverse resistivity $\eta_{EFF}$ which gave a magnetic diffusion time $\tau_B = \mu_0 L^2/(4\eta_{EFF}) = 0,5$ µs) of the same order as required. The driving mechanism for the current $\mathbf{J}_D$ was the (in the plasma's rest frame) time-changing magnetic field in the transition region. The most important observation for our purpose here concerns the current density $\mathbf{J}_D$, which was measured to be in a range where the relative velocity between the ions and the electrons ($U = v_{ez} - v_{iz}$) exceeds the ion thermal velocity by a factor $U/v_{thi} \approx 1,7 - 3,6$, where $v_{thi} = (3kT_i/m_i)^{1/2}$. In [5] the experimental parameter regime was extended from the single case studied in [3,4] to cover a factor 10 in density. Examples of the data are shown in Fig. 4. The diamagnetic cavity $\Delta B/B$ was in all cases maintained by about the same (normalized) diamagnetic current densities, with $U = v_{ez} - v_{iz}$ varying by $\pm 34\%$ around a mean value of $U/v_{thi} \approx 2,3$.

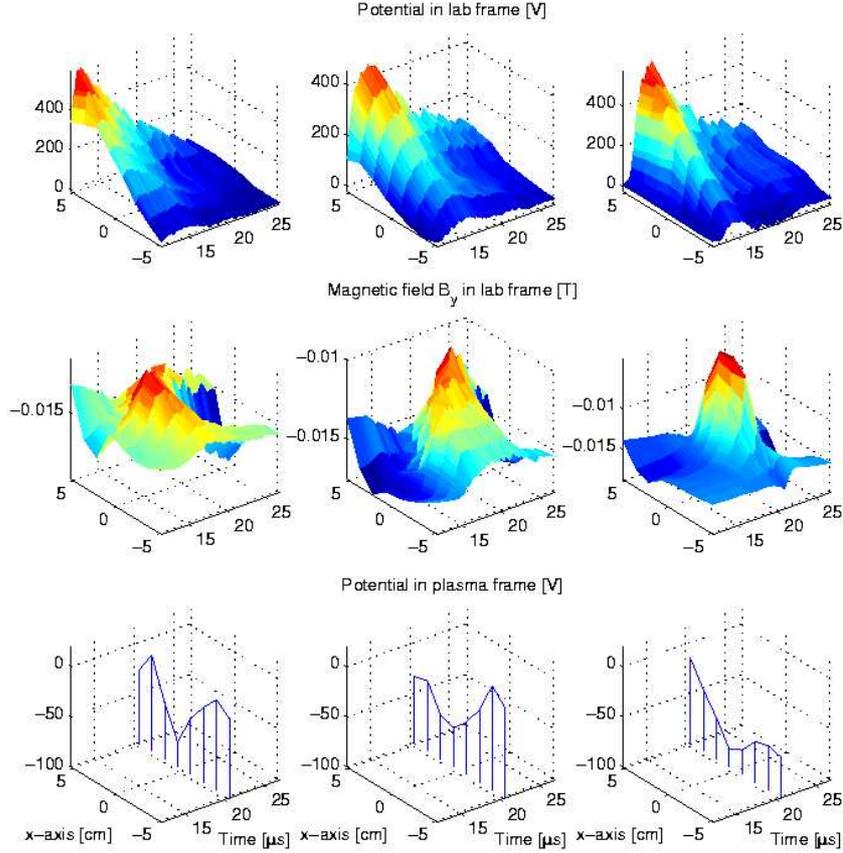

Fig. 4. Measurements with multi-probe arrays on the penetration of three plasmoids that vary a factor 10 in density, from [5]. The top plates show the plasma potential in the laboratory rest frame. The middle plates show the transverse magnetic field strength (notice that the diamagnetic cavities show up as peaks due to the negative sign of $B_y$). The bottom plates show the potential in the moving rest frame of the plasma. The plasma density and $\beta_k$ increases from left to right, and also the degree of magnetic expulsion in the centre (13%, 32%, and 74%).

The explanation proposed in [5] for this preferred value of $U/v_{thi}$ is basically the following, and is illustrated in Fig. 5. The wave amplitude, and thus the effective anomalous resistivity, should increase with the strength of the current $\mathbf{J}_D$ that drives the instability. Suppose that there is an increase of $\eta_{EFF}$ with $\mathbf{J}_D$ such as in Fig 5. This would give a self-organizing feedback loop that determines the gradient in the magnetic field during the magnetic diffusion process. The diamagnetic current density (and thus, through curl$\mathbf{B} = \mu_0 \mathbf{J}_D$, the gradient in field strength) will adjust to a value that depends on how strongly the process is driven. One example is the marginal stability (MS) analysis for weakly driven systems (see *e. g.* [11]). Such systems will stay close to the marginally stable state (A in Fig. 5). It was proposed in [5] that for strongly driven systems the diamagnetic current sheath should preferentially be located at the steep part of the curve, B-C-D in Fig. 5. Assume, for example, (at point D in Fig. 5) that the diffusion process is faster than the external driving mechanism, in the sense that the magnetic field profile flattens as a function of time. A flattening of the magnetic profile is equivalent to decreasing $\mathbf{J}_D$, going from D towards C in Fig. 5. This will decrease the value of $\eta_{EFF}$, reduce the magnetic diffusion speed, and counteract the flattening of the profile. If, on the other hand, the diffusion process (at point B in Fig. 5) is slow in the sense that that the magnetic gradient increases when the external magnetic field grows, then the reverse process would happen. In both cases, the profile tends to approach point C in Fig. 5. In the steep part of the curve a narrow range $\Delta \mathbf{J}_D$ corresponds to a large range $\Delta \eta_{EFF}$. Current densities lying rather close above some value B, where the curve begins to increase steeply, would be preferred in a large range in driving strengths of the diffusion. The steepness of the curve, as we have been able to estimate it from experimental data [5], is an order of magnitude larger than in Fig. 5: the increase of $\eta_{EFF}$ from the threshold (at A) to C is in the experiment a factor 200 – 300, while it in this figure is only a factor 25. A curve drawn to scale would in the range B-C-D be close to a vertical line.

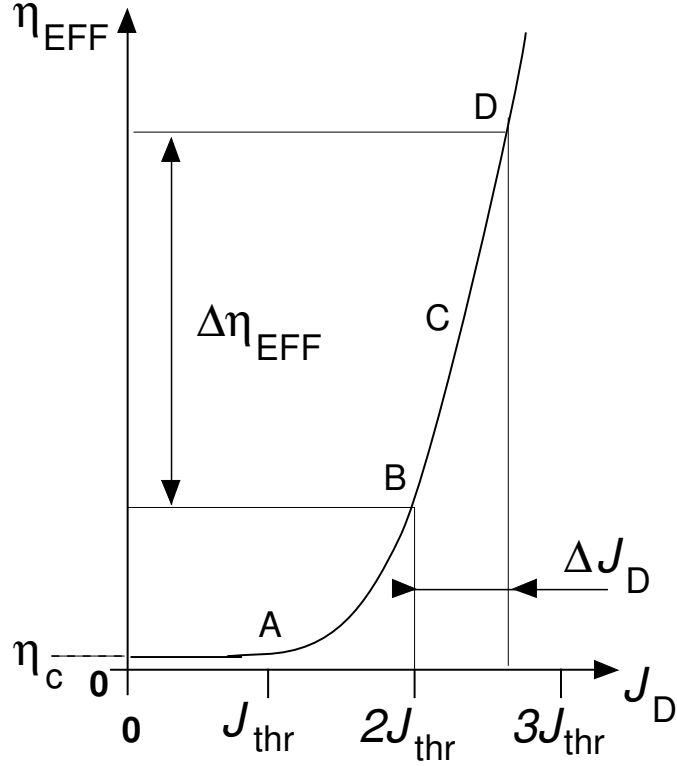

Fig. 5. A schematic diagram of the effective resistivity as function of the diamagnetic current density. The value $J_{thr}$ is the threshold for the MTSI, at current densities such that $U/v_{thi} \approx 1$.

The nonlinear magnetic diffusion (NL-MD) proposal in [4] can be expressed in three steps, essentially as follows: (1) for a plasmoid entering an abrupt magnetic barrier, the magnetic diffusion is very strongly driven. Only the fastest driven instability has time to grow and contribute to the diffusion; there should for this instability exist a functional relationship $\eta_{EFF}(\mathbf{J}_D)$ of the type shown in Fig. 5, which however needs to be properly normalized with respect to experimental parameters. (2) From the experimental data it was concluded that the operating instability is the MTSI, which has a threshold at diamagnetic currents such that $U/v_{thi} \approx 1$. In experiments over a range of 10 in plasma density, the diamagnetic current density was found to scale with density as $n_e^{-1}$, and was of such strength that $U/v_{thi} \approx 2,3$ with a variation of $\pm$ 34%. The quantity $U/v_{thi}$ is therefore proposed to give the proper normalization. (3) The diffusion rate (proportional to $\eta_{EFF}$ in Fig. 5) is proposed to vary so steeply with $\mathbf{J}_D$ at $U/v_{thi} \approx 2,3$ that this is a good approximation in a wide range of strongly driven magnetic diffusion. The assumption of nonlinear magnetic diffusion (NL-MD) is thus equivalent to assuming that the diamagnetic current density is given by

$$J_D = J_{NL-MD} = 2,3\, e n_e v_{ith}, \qquad (1)$$

where $v_{thi} = (3kT_i/m_i)^{1/2}$. Let us now test this against the four questions from Section 2 above.

| | $\beta_k$ | $\beta_{ith}$ | $w/r_{gi}$ | $[2,3(\beta_k\beta_{ith})^{1/2}]^{-1}$ | $\Delta B/B$ | Eq 2 | Eq 3 |
|---|---|---|---|---|---|---|---|
| Song *et al* [6] 50G | 1,35 | 0,034 | 3,7 | 2,0 | ≈ 100% | OK | - |
| Hurtig *et al* [3,4] | ≈ 1,6 | 0,03 | 0,4 | 2,0 | 13% | OK | 20% |
| Mishin *et al* [8] | ≈ 1 | 0,06 | 0,01 | 1,8 | 0,2% | OK | 0,5% |

Table I. Parameters for the experiments described in Section 2. The values for Mishin *et al* [8] have been estimated by us by scaling to the distance about 9m where $\beta_k \approx 1$. The two columns to the right are comparisons to the predictions from the NL-MD diffusion model. Eq. (2) predicts that the degree of magnetic expulsion depends on which is larger of $w/r_{gi}$ and $[2,3(\beta_k\beta_{ith})^{1/2}]^{-1}$. Eq. 3 gives an estimate of the expulsion $\Delta B/B$ for cases when $\Delta B/B \ll 1$.

### 3.1. High $\beta_k$ plasmoids : magnetic expulsion or self-polarization?

We first use Eq. (1) to determine when a plasmoid is in a state of marginal magnetic expulsion where $B = B_0$ outside, and $B = 0$ just in the middle of the plasmoid. If the plasma is elongated and has a width $w$, then Eq. (1) and Maxwell's Equation $\mathrm{curl}\mathbf{B} = \mu_0 \mathbf{J}_D$ gives [5] the relation $B_0 = \mu\, w J_D/2 = 2,3 e n_e v_{ith} \mu_0 w/2$. This can be rewritten using the dimensionless variables $w/r_{gi}$, $\beta_k$, and $\beta_{ith}$, where the first two have already been defined, $\beta_{ith} = W_{ith}/W_B$, and where $W_{ith}$ is the thermal ion energy density $n_i m_i v^2_{ith}/2$. Using these dimensionless variables, [5] found that the normalized beam width $w/r_{gi}$ determines if the magnetic field will be exposed from a plasmoid or not. The limit, marginal magnetic expulsion, is at

$$w/r_{gi} = [2,3(\beta_k\beta_{ith})^{1/2}]^{-1}. \qquad (2)$$

Wider beams should penetrate through magnetic expulsion, and much narrower beams by self-polarisation (and $\Delta B/B \ll 1$). Table I gives the parameters for the experiments shown in Fig. 2, leading to question 1 of section 2. These experiments were made at about the same $\beta_k$ value, but with different widths: Song *et al* [6] at 50 Gauss had a wide beam according to Eq. 2, and almost 100% magnetic expulsion from the whole plasma. Our own experiment [5] had a beam width about a factor 5 below the limit of Eq. 2, and was close to self polarization with $\Delta B/B = 13\%$. In the Porcupine experiment finally [8] the beam width was a factor 100 below the limit, and there was only 0,02 % magnetic expulsion. There is clearly good agreement between the data in Fig. 2 and Eq. 2.

## 3.2. Low $\beta_k$ plasmoids: how can the fast magnetic diffusion be energetically possible?

For low $\beta_k$ plasmoids, self-polarization is the only possibility to penetrate. However, there is always a residual degree of magnetic expulsion given by the diamagnetic current density of Eq. 1. Defining $\Delta B$ as the difference $(B_0 - B_c)$ between the centre and the surrounding, the NL-MD assumption of Eq. (1) corresponds [5] to a degree of expulsion

$$\Delta B/B_0 = 2{,}3(w/r_{gi})\,(\beta_k\beta_{ith})^{1/2} \,. \qquad (3)$$

A linear decrease from the edge to the centre corresponds to an average decrease of $\Delta B/2$ in magnetic field strength, and a change in magnetic energy which for small $\Delta B/B_0$ can be approximated as $\Delta W_B \approx B_0^2/2\mu_0 - (B_0 - \Delta B/2)^2/2\mu_0 \approx \Delta B\, B_0/2\mu_0$. This degree of magnetic expulsion can only be achieved if the directed kinetic energy is high enough, $W_K > \Delta W_B$. A little algebra gives the condition for low $\beta_k$ plasmoids to penetrate through self-polarization:

$$w/r_{gi} < (1/2{,}3)((\beta_k/\beta_{ith}) = (1/2{,}3)(v_0/v_{ith}). \qquad (4)$$

The physics can be understood as follows. When Eq. (4) is fulfilled, the beam energy is sufficient to create a magnetic cavity of the depth given by Eq. (3), with $\eta_{EFF}$ in the steep range B-C-D in Fig. 5. This sets the stage for further rapid magnetic diffusion into the plasmoid. The magnetic change $\Delta B$ does not grow any larger, and the kinetic energy in the plasma is sufficient to maintain it throughout the diffusion process. Notice that the plasma density does not enter into the energy condition of Eq. (4): there is, counter-intuitively, no lower limit to the plasma density motivated by the energy budget. Instead, the ion temperature (through $\beta_{ith}$) turns our to be important. The reason for this is straightforward. If the ion thermal speed is high, this will increase the threshold current density for the instability. In order to reach the NL-MD state of Eq. (1), a penetrating plasmoid will then need to push out a larger fraction of the magnetic field, which requires more energy.

## 3.3. What determines the low $\beta_k$, or electrostatic, limit for penetration?

In the experiment by Ishizuka *et al* [7], the electrostatic limit to penetration was found to be a factor of 10 above the theoretical limit which was $W_K = (m_i/m_e)^{1/2} W_E$. We propose that the limiting factor is the energy condition of Eq. (4). The limit found in their experiment [7] coincides excellently with Eq. (4), as shown in [5], figure 8. The description by Ishizuka *et al* [7] of the rejection mechanism is also consistent with this conclusion: as discussed in the companion paper [1, Fig. 8], there is an intimate connection between anomalous magnetic diffusion and anomalous electron transport across the magnetic field. Ishizuka *et al* [7] state that 'the electrons were stopped at the field boundary. The ions entered the field and created a positive sheath which caused the beam to be reflected rather than transmitted'. This is precisely what should be expected if the plasma stream did not have enough energy to drive a diamagnetic current of sufficient strength to (by means of the instability) allow the electrons to follow the ions into the transverse field region.

## 3.4. How can Lindberg's condition be reconciled with the experiment of Song et al [6]?

The question which is illustrated in Fig. 3 is how a plasmoid can be observed to have a width of 18 $r_{gi}$ (bottom panel) in view of Lindberg's condition that it shall be limited below 0,5 $r_{gi}$ (upper panel). The answer is that the plasmoid of Song *et al* [6] did not enter the transverse field in the manner envisaged by Lindberg, where individual ions have to climb a electrostatic and stationary potential hill $\Delta U = wE_P$ in the transition region (upper panel). Instead, we propose the following: our Eq. (2) indicates that the plasma beam in the experiment by Song *et al* first entered the transverse field by almost complete expulsion of the magnetic field close to the plasma source. In the following propagation the plasma beam expanded, and the magnetic field began to diffuse into it. In the measurements by Song *et al* [6] only the last stages of this diffusion was reported, from $\Delta B/B$ = 15% at 2,05m to 2% at 3.75m. The ions do not climb any potential $\Delta U = wE_P$ in such a process. They nevertheless lose energy, both to create the initial magnetic cavity and to drive the instability that enables the magnetic diffusion. These processes probably are involved in the observed velocity reduction from $7 \times 10^4$ m/s at the source to $2,8 \times 10^4$ m/s at 2,69 m. (We note that the lower ion speed gives a smaller gyro radius. The larger gyro radius obtained with the source velocity $7 \times 10^4$ m/s is not, however, enough to reconcile the experiment with Lindberg's condition. It is still violated by a factor of 14).

The relaxing of Lindberg's condition seems quite natural in this case, when the magnetic diffusion proceeds on a time scale much slower than the transition time into the transverse magnetic field. There is no obvious part of the device where the ions have a potential hill to climb. However, a closely related effect comes into play also for fast magnetic diffusion, when the plasma becomes self-polarized immediately in the transition. In this case the quasi-dc potential structures look just as in the upper panel of Fig. 3. The key quasi-dc electric field component is that along the plasma flow, in the *z* direction. The ions lose kinetic energy when they move against this field. However, when there is rapid magnetic diffusion in the transition region, there are also hf wave fluctuations there. A close look at how the wave $E_z$ field direction is correlated with the ion density [4, section VI d] reveals that the wave field systematically counteracts the quasi-dc electric field structure, and 'helps the ions to climb' to the high potential side. This relaxes Lindberg's condition, which is based on the ion energy budget. Wider beams than $r_{gi}/2$ are thus possible.

## 4. Discussion

We have found that one single, very simple, assumption gives a consistent explanation to a broad spectrum of observations for plasmoids that enter magnetic barriers. The assumption is that a diamagnetic current which is strongly driven will self-organize to a critical value where it corresponds to an electron-ion drift a few times the ion thermal speed, Eq. (1).

It should be noted that this model only applies to cases where the magnetic field changes so abruptly in the plasma's rest frame (with the transition time of the order of the ion gyro time) that only the very fastest instability, which we propose to be the MTSI, has time to grow and contribute to the magnetic diffusion mechanism. Also, we do not follow the process downstream. In this further motion, there will be continued diffusion of the magnetic field into a propagating plasmoid, the magnetic field change $\Delta B$ will decrease below that of equations (2) and (3), and the fast instability driven in the transition will eventually give place to slower instabilities, driven by lower diamagnetic current densities. This falls outside the scope of this work.

Some further results along the lines of the present work were reported in [5]. A parameter diagram was given, in which it can from basic parameters conveniently be assessed whether a given plasmoid will be rejected from a magnetic barrier, cross it by self polarization, or cross it by magnetic expulsion. It was also considered in [5] within what parameter range dense clumps in the solar wind, plasmoids, could enter the Earth's magnetosphere by the diffusive mechanism treated in this paper. Briefly summarized, solar wind plasmoids with width below about 35 km might penetrate the magnetosphere through self-polarization, while those of width larger than 350 km will need to penetrate mainly through magnetic expulsion. In the intermediate parameter range, 35-350 km, one has to look at the parameters for each individual case. Finally, a first step was taken towards a generalization of the NL-MD model, which here (Eq. 1) is based on the approximation of a constant diamagnetic current density. A functional relationship $\eta_{EFF}(\mathbf{J}_D, n_e, v_{ith}, T_e)$ was proposed that can be applied to any case with strongly driven magnetic diffusion, including time-dependent problems, and with space resolution of variable current densities $\mathbf{J}_D$. This model was applied to the transition across the magnetic barrier in our experiment and reproduced the experimental data quite acceptably, as shown in Fig. 6.

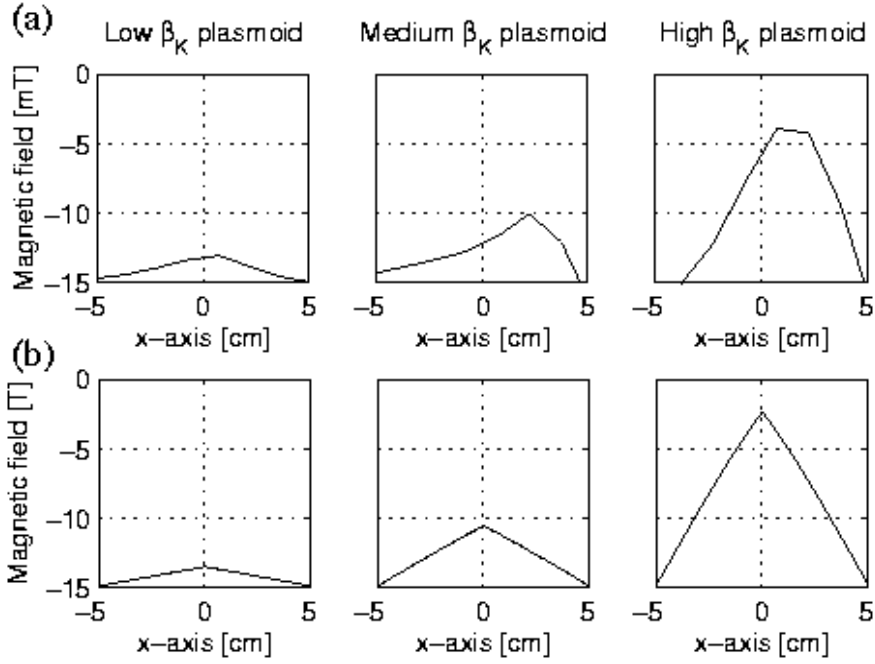

Fig. 6. A comparison between experimental data and model calculations (from [5]). The top row shows measured magnetic field profiles from the three experiments in Fig. 4. The bottom row shows calculations using the generalized formula for anomalous transverse resistivity $\eta_{EFF}(\mathbf{J}_D, n_e, v_{ith}, T_e)$ proposed in [5]. The asymmetry in the experimental data is a known feature of plasmoid injection, which has no counterpart in the model calculation.

## Acknowledgements

This work was supported by the Swedish Research Council.

## References

[1] Hurtig, T., Brenning, N., and Raadu, M. A., 2004, *The penetration of plasma clouds across magnetic boundaries: the role of high frequency oscillations for magnetic diffusion,* this issue.

[2] Hurtig, T., Brenning, N., and Raadu, M. A., 2003, *Three-dimensional particle-in-cell simulation with open boundaries applied to a plasma entering a curved magnetic field*, Phys. of Plasmas, **10**, 4291-4305.

[3] Hurtig, T., Brenning, N., and Raadu, M. A., 2004a, *The penetration of plasma clouds across magnetic boundaries : the role of high frequency oscillations*, accepted 2004 for Phys. of Plasmas.

[4] Hurtig, T., Brenning, N., and Raadu, M. A., 2004b *The role of high frequency oscillations in the penetration of plasma clouds across magnetic boundaries*, accepted 2004 for Phys. of Plasmas.

[5] Brenning, N., Hurtig, T., and Raadu, M. A., 2004*, Conditions for plasmoid penetration across magnetic barriers*, accepted 2004 for Phys. of Plasmas.

[6] Song, J. J., Wessel, F. J., Yur, G., Rahman, H. U., Rostoker, N., and White, R. S., 1990, *Fast magnetization of a high-to-low-beta plasma beam,* Phys. of Fluids **B2** (**10**), October 1990, 2482 – 2486.

[7] Ishizuka, H., and Robertson, S., 1982, *Propagation of an intense charge-neutralized ion beam transverse to a magnetic field,* Phys. Fluids **25** (**12**), Dec 1982, 2353 – 2358.

[8] Mishin, E. V., Kapitanov, V. I., and Treumann, R. A., 1986, *Anomalous diffusion across the magnetic field-plasma boundary : the Porcupine artificial plasma jet,* Journal of Geophysical Research**, 91**, 10183-10187.

[9] Peter, W., and Rostoker, N., 1982, *Theory of plasma injection into a magnetic field,* Phys. of Fluids B., **25**, 730-735.

[10] Lindberg, L., 1972, Astrophys. Space Sci **55**, 203.

[11] Manheiner, W. M., and Boris, J. P., 1972, Phys. Rev. Lett, **28**, 659-662.